\documentclass[prl, 
twocolumn,
showpacs ,aps]{revtex4}
\usepackage{times}
\usepackage{natbib}
\usepackage{graphicx}
\usepackage{amsmath}
\usepackage{bm}
\begin{document}
\title{A Single-Photon Server with Just One Atom}

\author{Markus Hijlkema}
\author{Bernhard Weber}
\author{Holger P. Specht}
\author{Simon C. Webster}
\author{Axel Kuhn$^{1}$}
\author{Gerhard Rempe}
\affiliation{Max-Planck-Institut f\"ur Quantenoptik, Hans-Kopfermann-Str. 1,  D-85748 Garching, Germany.\\ $^{1}$ Department of Physics, University of Oxford, Clarendon Laboratory, Parks Road, Oxford, OX1 3PU, United Kingdom }
\date{\today}
\begin{abstract}
Neutral atoms are ideal objects for the deterministic processing of quantum information. Entanglement operations have been performed by photon exchange\cite{Raimond01} or controlled collisions.\cite{Mandel03} Atom-photon interfaces were realized with single atoms in free space\cite{Darquie05,Volz06} or strongly coupled to an optical cavity.\cite{Kuhn02,McKeever04} A long standing challenge with neutral atoms, however, is to overcome the limited observation time. Without exception, quantum effects appeared only after ensemble averaging. Here we report on a single-photon source with one-and-only-one atom quasi permanently coupled to a high-finesse cavity. Quasi permanent refers to our ability to keep the atom long enough to, first, quantify the photon-emission statistics and, second, guarantee the subsequent performance as a single-photon server delivering up to 300,000 photons for up to 30 seconds. This is achieved by a unique combination of single-photon generation and atom cooling.\cite{Maunz04,Nussmann05,Murr06} Our scheme brings truly deterministic protocols of quantum information science with light and matter\cite{Cirac97,Pellizzari97,Browne03,Clark03-2,Duan04,Schoen05,Serafini06} within reach.
\end{abstract}

\maketitle


Deterministic single-photon sources are of prime importance in quantum information science.\cite{Monroe02} Such sources have been realized with neutral atoms, embedded molecules, trapped ions, quantum dots, and defect centres.\cite{Grangier04} All these sources are suitable for applications where the indivisibility of the emitted light pulses is essential. For quantum computing or quantum networking, the emitted photons must also be indistinguishable. Such photons have so far only been produced with quantum dots\cite{Santori02} and atoms.\cite{Legero04,Beugnon06} Another requirement is a high efficiency. This is hard to obtain in free space, as the light collecting lens covers only a fraction of the full $4\pi$ solid angle. The efficiency can be boosted by strongly coupling the radiating object to an optical microcavity, as has been achieved with atoms\cite{Kuhn02,McKeever04} and quantum dots.\cite{Press06} An additional advantage of the cavity is that a vacuum-stimulated Raman adiabatic passage can be driven in a multilevel atom.\cite{Hennrich00} In this way amplitude,\cite{Kuhn02,Keller04} frequency,\cite{Legero04} and polarization\cite{Wilk07} of the photon can be controlled. It should also make possible to combine partial photon production with internal atomic rotations for the construction of entangled photon states such as W and GHZ states.\cite{Schoen05} 
\begin{figure}[h]
\centerline{\includegraphics[width=8.0cm]{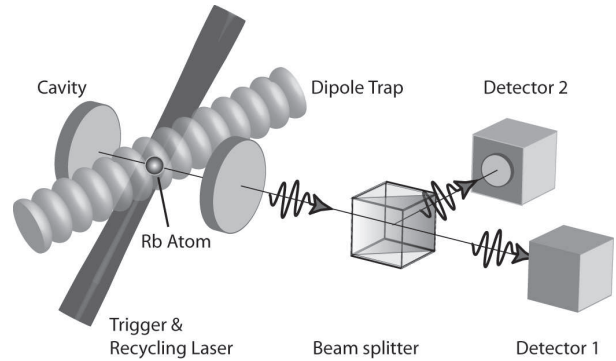}}

 \caption{ \small \textbf{Schematic of the apparatus.} A single $^{85}$Rb atom is trapped in a high-finesse optical microcavity by means of a two-dimensional optical lattice. Confinement along the cavity axis and a direction perpendicular to it is achieved with a weak cavity-stabilization laser and a strong retroreflected dipole laser, respectively. Confinement along the third direction results mainly from the small beam waist of the dipole laser. The atom-cavity system is excited by a sequence of laser pulses incident under an angle of $45^\circ$ to the dipole laser and perpendicular to the cavity axis. Single photons emitted from the system are detected by two avalanche photodiodes in Hanbury Brown \& Twiss configuration. For simplicity, details of the setup such as the set of prisms and interference filters in front of the detectors are not shown.} \label{fig:Setup}
\end{figure}

All these demands together have so far only been achieved with atoms in high-finesse microcavities. A reason is that neutral atoms are largely immune to perturbations, such as electric patch fields close to dielectric mirrors. However, atomic systems have always suffered from a fast atom loss. We have now implemented a cavity-based scheme, see figure \ref{fig:Setup}, with a dipole laser for trapping, a trigger laser for photon generation and a recycling laser for repumping, monitoring and cooling the atom.\cite{Nussmann05,Murr06} The scheme combines high photon-generation efficiency and long trapping times. The most remarkable features are, first, that the single-photon stream is specified by its intensity correlation function evaluated in realtime during a short time interval after system preparation and, second, that its subsequent performance is guaranteed by monitoring the atom without perturbing the single-photon stream. This makes our single-photon source a truly useful quantum device operating with not more than just one single atom. 


The main parts of the apparatus are described elsewhere,\cite{Nussmann05,Nussmann05-2} but changes were made to allow for single-photon generation and detection in combination with atom cooling. In short, $^{85}$Rb atoms are collected from a background vapour in a magneto-optical trap, loaded into a running-wave dipole trap (wavelength 1032 nm) and transferred into the optical high-finesse cavity. Upon arrival, a few atoms are captured by switching the geometry of the dipole trap to a standing wave and turning on 780 nm lasers perpendicular to the cavity axis for three-dimensional cavity cooling.\cite{Nussmann05} The cavity has length 0.5 mm, mode waist $29 {\mu}$m and finesse $3\times10^{4}$. The relevant parameters are $(g, \kappa, \gamma)\,=\,2\pi\times(5, 5, 3)\,$MHz, with $g$ the maximum atom-cavity coupling constant on the $F=2\to{F'}=3$ transition between the atomic $5S_{1/2}$ ground and $5P_{3/2}$ excited state, $\kappa$ the cavity-field decay rate and $\gamma$ the atomic dipole decay rate. One of the cavity mirrors has a 50 times higher transmittance than the other. Photons scattered into the cavity by the trapped atom and emitted through this output mirror are spectrally and spatially filtered from the light of a cavity-stabilization laser at 785 nm and stray light, respectively, by means of cascaded glass prisms in combination with interference filters and pinholes. Finally, the photons are counted by two avalanche photodiodes. The combined background count rate due to stray light and dark counts is 84 Hz.

\begin{figure}
\centerline{\includegraphics[width=8.0cm,clip=true]{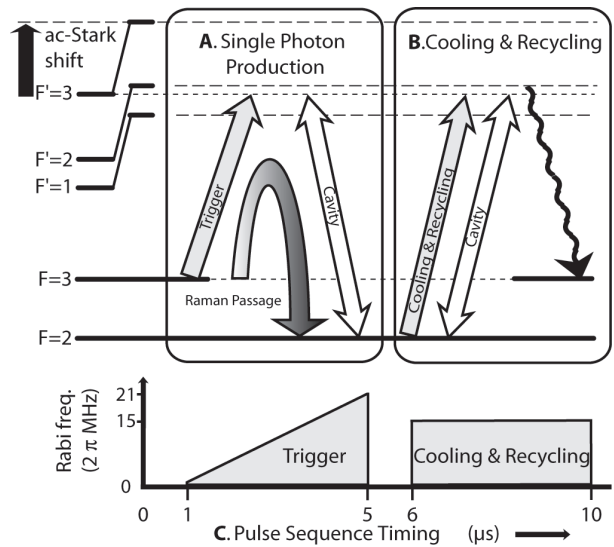}} 

 \caption{\small \textbf{ Simplified energy level diagram.} During single-photon production the $^{85}$Rb atom is excited by a $4{\mu}s$ long pulse of a trigger laser resonant with the $F=3\to{F'}=3$ transition between the $5S_{1/2}$ ground and $5P_{3/2}$ excited state. With the cavity resonant with the $F=2\to{F'}=3$ transition, this drives a vacuum-stimulated Raman adiabatic passage from $F=3$ to $F=2$, leading to the emission of a photon from the cavity. $1{\mu}s$ later, a $4{\mu}s$ long laser pulse is used to recycle the atom back into the $F=3$ state. This laser is resonant with the cavity and the $F=2\to{F'}=3$ transition. As the presence of the trapping potential leads to a dynamic energy-level shift (the ac-Stark shift) of $\Delta_S\approx2\pi\times70$ MHz, 7 MHz larger than the atomic hyperfine splitting between the ${F'}=2$ and ${F'}=3$ states, all lasers and the cavity are red detuned from the atomic transitions from the ground states to the $F'=2$ excited state by a few MHz. This results in a situation where the atom can cycle a few times, scattering photons into the cavity, before falling back into the $F=3$ state. This cycling cools the atom. }\label{fig:Scheme}
\end{figure}

Figure \ref{fig:Scheme} depicts the three-level system that enables single-photon production and atom cooling. Starting with a single atom in the $F=3$ ground state, a trigger pulse together with the cavity drives a vacuum-stimulated Raman adiabatic passage\cite{Kuhn99,Hennrich00} into the $F=2$ ground state (Fig. \ref{fig:Scheme}A). This generates a single photon that is emitted from the cavity. Next, the atom is pumped back to the initial $F=3$ state with a recycling laser resonant with the cavity (Fig. \ref{fig:Scheme}B). During this recycling process, the atom can scatter many photons into the cavity. To understand the scattering process in more detail, the $k_B\times1.5$ mK deep dipole trap has to be taken into account ($k_B$ is Boltzmann's constant). The trap shifts the atomic resonances by $\Delta_S\approx2\pi\times70$ MHz, the dynamic Stark shift. As displayed in Fig. \ref{fig:Scheme}A~and~B, all lasers and the cavity are not resonant with the atom. This has little consequences for the photon production in the Raman process as both ground states experience the same shift. For the recycling laser, however, it creates a situation similar to one described earlier, with a strong Sisyphus-like cooling force.\cite{Nussmann05,Murr06} This occurs even though the employed atomic transition is not closed so that cooling takes place only as long as the atom cycles between the ground and excited state. Note that full three-dimensional cavity cooling is used only for the initial trapping of the atom. Here both the trigger laser (acting now as a repumper) and the recycling laser are turned on continuously. This can keep the atom in the cavity for up to a minute. 

While applying the photon-production and recycling pulses, atoms stay in the cavity for 10.3(1) s on average, as determined from 526 experimental runs. This is about twice as long as in the dark dipole-force trap. Taking into account that initially several atoms are trapped and that one has to wait until all but one atom have escaped the cavity, single atoms are available for single-photon production for 8.3(2) s on average. This gives 4379 s of data with $4.2\times10^6$ detection events. For a trigger rate of 100 kHz, the overall photon generation, propagation and detection probability then amounts to 0.93\%. This includes 50\% cavity absorption loss mainly due to a mirror defect, 52\% propagation loss from the cavity to the detectors and 44\% quantum efficiency of the detectors. The photon-generation probability is therefore 9\%. The finite efficiency is attributed to the large number of Zeeman states, some of them exhibiting a small atom-cavity coupling constant. 

We calculate the cross-correlation of the recorded photon stream binned over $4 \mu s$ long intervals corresponding to the trigger pulses (discarding detection events outside the trigger pulse). Summed up over the 526 runs this gives 1.2$\times10^4$ correlations on average for time differences corresponding to different trigger pulses. This gives 22 correlations per atom per bin over the average lifetime. The measured antibunching visibility is 94.0\%. 

\begin{figure}
\centerline{\includegraphics[width=8.0cm]{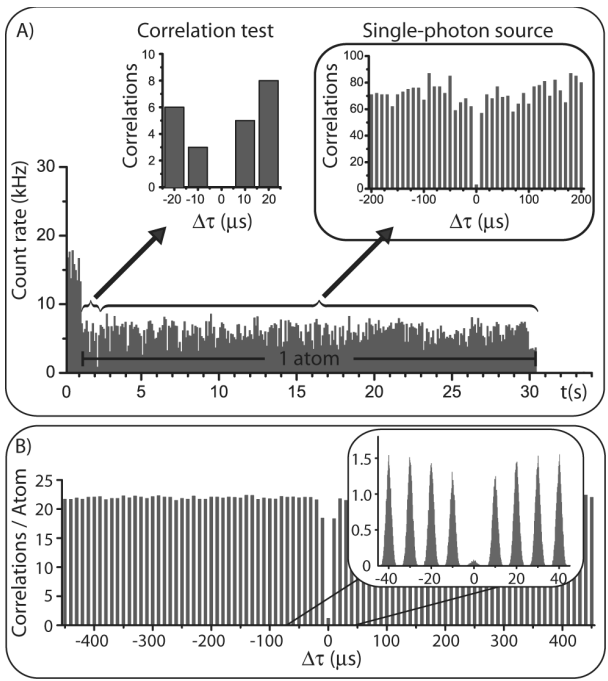}}

 \caption{ \small \textbf{(A) Deterministic single-photon source with just one atom.} The light emitted from the cavity during single-photon production and recycling is plotted versus time for one experimental run. Initially, a few atoms are trapped in the cavity. After 1 s the average count rate drops to the single-atom level. To verify that exactly one atom is trapped, we calculate the cross correlation of the photons recorded by the two detectors during trigger pulses for the next 1.5 s. A single atom manifests itself by the absence of coincidences for zero detection-time difference, $\Delta \tau$=0, in the correlation function, as shown in the upper left inset. The upper right inset displays the correlation function obtained from the remaining 28 s long single-photon stream, containing $\sim2.8\times10^5$ single photons, of which we detect 11\%. Antibunching has a visibility of 95.8\%. \textbf{(B) Average behaviour}. Correlation function averaged over those 454 single-atom runs which passed the qualification procedure described in the text. Antibunching has a visibility of 94.6\%, limited only by background counts. The similarity of the single-atom trace shown in (A) compared with the average behaviour demonstrates the deterministic character of our source. The inset displays the averaged correlation function with a time resolution of 200 ns.} \label{fig:Results}
\end{figure}

The large number of correlations observed per atom and the large visibility of the antibunching suggests the following measurement protocol for single-atom operation, see figure \ref{fig:Results}: First, the system is initialized by trapping a few atoms and monitoring the light level emitted from the cavity during recycling pulses. When this level reaches a value expected for one atom (4 photons/ms on average), the photons emitted during trigger pulses within the next 1.5 s are recorded and the cross-correlation of the (binned) photon stream is calculated, see Fig.\ref{fig:Results}A, left inset. Next, the data are tested against the selection rule that the average number of correlations for non-zero time differences must exceed 1.5 (to make sure that at least one atom is trapped, leading to 4 correlations on average) and that correlations at zero time difference must not exceed 30\% of this average (to make sure that not more than one atom is trapped). We find that 86\% of all 526 runs pass such a test. This leaves 454 runs with 3774 s of true single-photon data with on average $1.0\times10^{4}$ correlations at non-zero time difference and 534 coincidences at zero time difference. From our background count rate we would expect 587(24) coincidences of a photon click with a background click. The measured antibunching with its visibility of 94.6\% is therefore entirely limited by dark counts and stray light.

A single-atom source that passed the described selection procedure therefore emits a high-quality stream of single photons. A user of the photons can be notified and the photons redirected as needed. While this is done, the presence of the atom is monitored by detecting the light emitted during the recycling pulses. A loss of the atom manifests itself by the absence of scattered photons which can be detected within $\sim$30 ms with 98\% probability. In the right inset of figure \ref{fig:Results} the photon correlation function is plotted for those photons that would have been sent to the user. For comparison, the correlation function for all 454 runs is depicted in figure \ref{fig:Results}. The correlation signal obtained for one-and-the-same atom clearly displays the antibunching that is otherwise observed only after averaging over an ensemble of single atoms. Note that in contrast to all previous single-atom experiments, the single-atom nature of our system is obtained from a nonclassical correlation signal, not a classical average of the emitted photon stream. This unambiguously discriminates a single atom from several atoms. 

In conclusion, our atom-cavity system has progressed from a proof-of-principle single-photon source to a useful device whose performance is specified during operation. The quasi-permanent availability of exactly one atom, the high efficiency of photon production in a well-defined light mode, and the large duty cycle of the whole measurement sequence paves the way for deterministic atom-photon and atom-atom entanglement experiments such as a test of Bell's inequality with distant atoms. 

\footnotesize\noindent\textbf{Acknowledgements} This work was supported by the Deutsche Forschungsgemeinschaft [SFB 631 and Research Unit 635] and the European Union [IST (SCALA) and IHP (CONQUEST) programs].

\end{document}